\documentclass[showpacs,twocolumn,prd]{revtex4-1}

\usepackage{slashed}
\usepackage{footnote}
\usepackage{mathrsfs}
\usepackage{amsfonts}
\usepackage{amsmath}
\usepackage{amssymb}
\usepackage{revsymb}
\usepackage{graphicx}
\usepackage{mathrsfs}
\usepackage{bm}
\usepackage{bbm}
\usepackage{psfrag}
\usepackage{color}

\usepackage[normalem]{ulem}
\usepackage{natbib}

\usepackage[usenames,dvipsnames]{xcolor}
\usepackage{multirow,tabularx}
\usepackage{xcolor,pifont}
\newcommand*\colourcheck[1]{%
  \expandafter\newcommand\csname #1check\endcsname{\textcolor{#1}{\ding{52}}}%
}

\newcommand{\be}{\begin{equation}} 
\newcommand{\ee}{\end{equation}} 
\newcommand{\bea}{\begin{eqnarray}} 
\newcommand{\eea}{\end{eqnarray}} 

\newcommand{\eps}{\varepsilon}

\newcommand{\trm}[1]{\textrm{#1}}

\newcommand{\figref}[1]{Fig. \ref{#1}}

\newcommand{\eqnref}[1]{Eq.\,(\ref{#1})}
\newcommand{\eqnrefs}[2]{Eqs.~(\ref{#1}) and (\ref{#2})}

\newcommand{\tr}{\trm{tr}\,}

\newcommand{\psibar}{\bar{\psi}}
\newcommand{\x}[1]{\stackrel{\tsf{x}}{#1}}
\newcommand{\tsf}[1]{\textsf{#1}}

\newcommand{\vkap}{\varkappa}
\newcommand{\vphi}{\varphi}

\newcommand{\nn}{\nonumber}
\newcommand{\J}{\trm{J}}

\newcommand{\pin}{p_{\trm{in}}}
\newcommand{\non}{\nonumber}

\definecolor{oorange}{rgb}{1.0,0.5,0.0}

\usepackage{amssymb}

\begin{document}

\title{Light scalars:  coherent nonlinear Thomson scattering and detection}
 
\author{B.~M.~Dillon}
\author{B.~King}
\affiliation{Centre for Mathematical Sciences, University of Plymouth, Plymouth, PL4 8AA, United 
Kingdom}
\email{barry.dillon@plymouth.ac.uk}
\email{b.king@plymouth.ac.uk}

\date{\today}
\begin{abstract}
Several theories of beyond-the-standard-model physics predict light scalars that couple to fermions. By extending classical electrodynamics to include an electron-scalar coupling, we calculate the nonlinear Thomson scattering of light scalars in the collision of an electron with a monochromatic electromagnetic background. 
In doing so, we identify the classical electron-scalar current, which allows for straightforward inclusion of the process in laser-plasma particle-in-cell simulations.
Scattering of pseudoscalar particles is found to vanish in the classical (or, equivalently, the low-lightfront-momentum) limit. 
When electrons \emph{co-propagate} with the laser pulse, we demonstrate that coherence effects in the production of light scalar particles can greatly enhance the signal for sub-eV scalars.
When the electron beams \emph{counter-propagate} with the laser pulse, we demonstrate that experiments can probe larger scalar masses due to the larger momentum transfer in the collisions.
We then discuss a possible lab-based experimental set-up to detect this scalar signal which is similar to light-shining-through-the-wall experiments.  
Using existing experimental facilities as benchmarks, we calculate projected exclusion bounds on the couplings of light scalars in such experiments. 
\end{abstract}
\maketitle

\section{Introduction}

\noindent There are many candidates for light beyond-the-standard-model particles, some of which can couple directly to spin-1/2 fermions and can therefore be emitted in electron-laser interactions. 
One such candidate is the axion, a spin-0 pseudoscalar particle predicted by the Peccei-Quinn solution to the strong CP problem \cite{Peccei:1977hh}.
However, other light candidates include scalar particles, dark photons \cite{ahlers07}, or even milli-charged particles \cite{raffelt88}.
Collectively these particles are referred to as Axion-Like-Particles (ALPs).
Many experimental searches for ALPs have already been performed both using lab-based and astrophysical sources (see \cite{Graham:2015ouw,Irastorza:2018dyq} for recent reviews).

In this paper, we build on previous works \cite{king18c,king18b}, in which we studied ALP production in laser-electron interactions, to detail how the coherent emission of scalar ALPs from electron bunches in laser interactions could allow one to obtain a competitive bound on the coupling of scalar ALPs to electrons and photons.
We focus on scalar rather than pseudoscalar ALPs because, as we show in this paper, the latter have a suppressed production rate in the low-energy, coherent, limit.
Scalar-ALPs arise in many beyond-the-standard-model scenarios, for example pseudo-Goldstone bosons of spontaneously broken global symmetries, or as dilaton fields from the spontaneous breaking of scale symmetry. Massive scalars also occur in cosmological contexts, for example from quintessence fields \cite{tsujikawa13} or the inflaton \cite{lyth91}.

The coupling of scalar ALPs to the photon is already constrained by fifth-force experiments \cite{Dupays:2006dp}, since the scalar-photon coupling induces a coupling between the scalar and the proton that mediates a long-range spin-independent non-Newtonian force. (The prospect of using intense laser pulses to probe photon-ALP coupling has also been explored in the literature \cite{gies09,*villalba-chavez14,*villalba-chavez15}.)
The bounds obtained from fifth force experiments are typically much stronger than those from lab-based experiments such as Light-Shining-through-Wall (LSW) set-ups (see \cite{Redondo:2010dp} for a review on LSW experiments).
However not only are these bounds only applicable to sub-eV scalars, as discussed in \cite{Dupays:2006dp}, the fifth force bounds can be much weaker when one considers effects that modify the form factor coupling the scalar to the photon.
The same argument suggests that bounds from astrophysical sources, such as CAST \cite{CAST17,Barth:2013sma}, could also be much weaker than those quoted when mechanisms are at play that reduce the rate of ALP production either as a whole or in a particular energy range \cite{Masso:2005ym,Jain:2005nh,Jaeckel:2006id,Masso:2006gc}.
A major motivation for these works was the apparent signal at the PVLAS experiment \cite{Zavattini:2005tm}, which contradicted existing bounds using astrophysical sources and has since vanished \cite{Zavattini:2007ee}.
However given the existence of scenarios in which bounds from fifth-force experiments and astrophysical experiments can be evaded, the need for lab-based searches for light ALPs is apparent.
Therefore in this paper we propose a new mechanism through which the coupling between scalar ALPs and the electron can be probed to high accuracy in a lab-based environment. We consider the probing of scalar ALPs with masses up to $\mathcal{O}(100)\,$eV.

The experimental set-up that we propose will consist of an electron bunch, which we can treat as a plasma, colliding with a laser pulse.
When simulating interactions between plasmas and intense laser backgrounds, one typically splits processes into two groups: 
\emph{incoherent}, single (dressed) particle processes which occur at wavelengths much smaller than the electron spacings in the bunch, and 
\emph{coherent} processes which proceed at lower energies with wavelengths of the order of the electron spacings in the bunch.
If a process is coherent over the entire bunch i.e. the wavelength of the emitted particle is longer than the bunch length, then the rate scales with the number of electrons in the bunch squared.
These processes are simulated using traditional particle-in-cell (PIC) simulation techniques \cite{ruhl10,*king13a,*gonoskov15,*vranic16,*tamburini17}. 
Due to the large possible enhancement in the yield of coherent processes and the impact that can bring to ALP searches, in the current paper, we focus on the calculation of scalar emission from an electron in the low-lightfront-momentum (classical) regime.

The paper begins in Sec. II with a discussion on the classical calculation of scalar emission from an electron bunch in a continuous-wave laser (monochromatic electromagnetic background), and we comment on the inclusion of such processes in PIC code simulations. In Sec. III, the classical result is compared to the classical limit (equivalently: small incoming lightfront momentum) of the full QED calculation of the process. In Sec. IV, we investigate coherent emission of ALPs from an electron bunch interacting with a laser. In Sec. V we discuss the experimental prospects for scalar ALP production and in Sec. VI derive and present exclusion plots for the result of such an experiment.  In Sec. VII we conclude and in App. A we add a note explaining the suppression in the low-energy limit of pseudoscalar production.

\section{Nonlinear Thomson scattering of scalars in a monochromatic background} \label{NLTcalc}

\noindent The interaction of an electron and a scalar field, $\phi$, in a laser pulse background can be described using the following Lagrangian density (unless otherwise stated, we have set $\hbar=c=1$)
\bea
\mathcal{L} = \mathcal{L}^{0}_{\phi} + \mathcal{L}^{0}_{\tsf{SFQED}} + \mathcal{L}^{I}_{\phi} + \mathcal{L}^{I}_{\tsf{SFQED}} + \mathcal{L}^{I}_{\phi\gamma\gamma}, \label{eqn:L}
\eea
where
\begin{align}
\mathcal{L}^{0}_{\phi} &= \frac{1}{2} \left[(\partial \phi)^{2} - m_{\phi}^{2}\phi^{2}\right] 	\non\\
 \mathcal{L}^{0}_{\tsf{SFQED}} &= \bar{\psi}\left[i\left(\slashed{\partial}+ie\slashed{A}_{\trm{laser}}\right)-m_e\right]\psi - \tr F^{2}/4	
 \end{align}
are the free-field real scalar and dressed Strong-Field QED (SFQED) parts, respectively, with the scalar being neutral under electromagnetism.
The interaction terms are 
\begin{align}
\mathcal{L}^{I}_{\phi}&=-g_{\phi e} \phi \bar{\psi} \psi	 \non\\
\mathcal{L}^{I}_{\tsf{SFQED}}&= -e \bar{\psi}\slashed{A}_{\gamma}\psi \non\\
\mathcal{L}^{I}_{\phi\gamma\gamma}&=-g_{\phi\gamma\gamma}\phi  F^{\mu\nu}_{B}F_{B\,\mu\nu}
\end{align}
where $e>0$ is the charge of a positron, $g_{\phi e}$ the scalar-electron coupling. The dimension-five interaction, $\mathcal{L}^{I}_{\phi\gamma\gamma}$, will become relevant when discussing regeneration of the scalars into photons in a magnetic field in the detection region, discussed in Sec. \ref{ExpProspects}.
We have made the split $F \to F_{\trm{laser}} + F_{B} + F_{\gamma}$, into i) a (classical) laser field (to generate scalars) and ii) a (classical) magnetic field (to regenerate photons) and iii) a (quantum) radiated field, respectively (the classical-quantum split is standard in SFQED -- for reviews, see \cite{ritus85,*dipiazza12,*narozhny15,*king15a}). (Labels on the vector potential, $A$, reflect the corresponding field.)  The generation and regeneration regions are distinct so that $F^{\mu\nu}_B\,F^{\rho\sigma}_{\trm{laser}}=0$.
In SFQED, the interaction between the laser background and the electron is included exactly by solving for the particle dynamics exactly in a plane-wave electromagnetic (EM) background of phase, $\vphi = \vkap\cdot x$, and wavevector $\vkap$ satisfying $\vkap\cdot\vkap =0$. In the quantum theory, this amounts to using the Volkov solution to the Dirac equation \cite{volkov35} whereas in the classical theory, this means solving the Lorentz equation \cite{landauC}.

Two useful parameters for quantifying the size and nature of SFQED processes in plane waves are the \emph{classical nonlinearity parameter}, $\xi$, and the \emph{quantum nonlinearity parameter}, $\chi$.
The classical nonlinearity parameter can be written as \cite{ilderton09} $\xi^{2} = e^{2} \langle p\cdot T(\vphi)\cdot p \rangle_{\vphi}/m_e^{2} (\vkap\cdot p)^{2}$, where $T$ is the laser pulse stress-energy tensor, $p$ the electron momenta and $\langle \cdot \rangle_{\vphi}$ an average over field phase. $\xi$ is then equal to the work done by the laser pulse on an electron over the electron's Compton wavelength, divided by the energy of a photon and hence quantifies the average number of photons from the laser background that interact with a single electron. 
Our analysis will take into account arbitrary values of $\xi$, but we expect any likely first laser-plasma-ALP experiment will take place at $\xi \ll 1$, that is, where interaction between the electron and the laser can be assumed to be \emph{perturbative}. 
The quantum nonlinearity parameter for a particle of momentum $p$ can be written as $\chi_{p} = \xi \eta_{p}$ in a plane wave, where $\eta_{p} = \vkap\cdot p / m_e^{2}$. It is so called because $\chi_{p} \propto \hbar$ and hence disappears in the \emph{classical} limit of $\hbar \to 0$. In this work, we will use the lighfront momentum variables $\eta_{p}$, $\eta_{k}$ (which are also $\propto \hbar$), to quantify the size of quantum effects. (We will typically take $\xi=\mathcal{O}(0.1)\ldots \mathcal{O}(10)$).

As mentioned in the introduction, one of our interests lies in the \emph{coherent} emission of scalar particles. For this to happen over an entire electron bunch, the scalar wavelength should be much longer than the bunch length \cite{berryman96} and in this respect, we are interested in the limit $\eta_{k}\to 0$. 
As the magnitude of $\chi_{k}$ is limited by $\chi_{p}$ in the Compton case, the coherence effects are important in the limit $\eta_{p} \to 0$. 
This corresponds to neglecting electron recoil from photon emission and hence is synonymous with the classical limit.

We wish to calculate the process $e^{\ast}  \to e^{\ast} + \phi$, where $e^{\ast}$ indicates an electron ``dressed'' in the laser pulse background, in the classical (low-lightfront-momentum) regime. 
Due to the smallness of the electron-scalar coupling, it is clear that the more probable process is that of nonlinear Compton scattering, $e^{\ast}  \to e^{\ast} + \gamma$. 
However, we can neglect the effect this has on the electron trajectory (i.e. radiation reaction), if we assume $\alpha \xi \chi \ll 1$, and $\chi \ll 1$ \cite{dipiazza10,ilderton14}, where $\alpha =e^{2}/4\pi$. 
Then from the Lagrangian \eqnref{eqn:L}, in the generation region (where $F_{B}=0$), we find:
\bea
\left(\Box + m^{2}_{\phi}\right)\phi &=& -g_{\phi e}\psibar\psi   \nn \\
\Box A^{\mu}_{\gamma} &=& e \psibar \gamma^{\mu} \psi   \nn \\
\left[i\left(\slashed{\partial}+ie\slashed{A}_{\trm{laser}}\right) - m_e\right]\psi &=& 0\nn  \\
\psibar\left[-i\left(\slashed{\partial}-ie\slashed{A}_{\trm{laser}}\right) - m_e\right] &=& 0,  \label{eqn:sys1}
\eea
where we have assumed the Lorentz gauge: $\partial \cdot A = 0$. Let $\psi = \psi^{(0)} + e \psi^{(1)} + \ldots$ be a perturbative ansatz in the electron-photon coupling, and let $g_{\phi e} \ll |e|$, then these equations can be decoupled to give:
\bea
\left(\Box + m^{2}_{\phi}\right)\phi &=& j; \qquad j = -g_{\phi e}\psibar^{(0)}\psi^{(0)}, \label{eqn:sys1einfach}
\eea
where $\psi^{(0)}$ solves the two Dirac equations in \eqnref{eqn:sys1} exactly in whatever plane wave potential is chosen to describe the laser pulse. 

As the scalar current $j$ is simply proportional to the number density of the electrons, we make the correspondence with a classical number density $n(x)$ using:
\bea
 g_{\phi e}\psibar^{(0)}(x)\psi^{(0)}(x) \to 2g_{\phi e} n(x). \label{eqn:swap}
\eea
The extra factor $2$ in \eqnref{eqn:swap} comes from the fact that:
\bea
\langle \tr\,\psibar\psi \rangle_{\tsf{spin}} = (2\,m_e)^{2},
\eea
where $\langle \cdot \rangle_{\tsf{spin}}$ refers to an average over initial electron spins.
The factor $2$ in \eqnref{eqn:swap} therefore takes into account the spin-sum of standard QED which has no meaning in the classical calculation. 

Having made the identification in \eqnref{eqn:swap}, we now proceed to solve the \emph{classical} version of \eqnref{eqn:sys1einfach} using 
\bea
\phi(x') = \int D(x'-x) j(x) d^{4}x,
\eea
with $(\Box + m^{2})D(x'-x) = \delta^{(4)}(x'-x)$.
To demonstrate our results, we calculate the ALP spectrum produced in the case of a circularly-polarised monochromatic background \cite{king18b}:
\bea
a^{\mu} = m\xi \left(\eps^{\mu} \cos \vphi + \beta^{\mu} \sin \vphi\right),
\eea
where $\eps \cdot \beta = \eps\cdot \vkap =\beta \cdot \vkap = 0$, $\eps\cdot \eps = \beta\cdot \beta = -1$ and $a = eA$ is the scaled vector potential.
The yield of scalars, $N_{\phi}$, from a classical source is equal to \cite{Peskin:1995ev}:
\bea\label{yield1}
N_\phi = \int \frac{d^{3}k}{(2\pi)^{3}}\frac{1}{2k^{0}} |\widetilde{\jmath}(k)|^{2},
\eea
where $\widetilde{\jmath}(k) = \int d^{4}x ~\mbox{e}^{i\,k\cdot x}\,j(x)$ is the Fourier-transformed current. We then write the classical scalar current by analogy with the EM current \cite{jackson99}:
\bea
j\left(x'^{\mu}\right) = -g_{\phi e} \int d\mathfrak{t}~\delta^{(4)}\left[x'^{\mu}-x^{\mu}(\mathfrak{t})\right],\label{eqn:swap2}
\eea
and $\mathfrak{t}$ is the proper time. The electron's position in a plane wave can be solved for exactly \cite{landauC}:
\bea
\frac{x^{\mu}(\vphi)}{m_e}\!=\! \int^{\vphi}_{-\infty} \!\!\frac{\pin^{\mu}-a^{\mu}(\phi)}{\vkap \cdot \pin} + \vkap^{\mu} \left(\frac{-a^{2}(\phi) + 2\pin \cdot a(\phi)}{2(\pin\cdot \vkap)^{2}}\right)\!d\phi \nn \\
\eea
where $\pin$ is the electron's asymptotic momentum, before it meets the laser pulse. The calculation proceeds in a very similar manner to the quantum case \cite{king18b}, and we arrive at a rate $\tsf{R}^{e\to\phi} = \sum_{s \geq s_{0}^\phi}\tsf{R}^{e\to\phi}_{s}$
\bea
\tsf{R}^{e\to\phi}_{s} &=&  \frac{g_{\phi e}^{2}}{4\pi\eta_{p}} \int_{t^{-}_{s}}^{t^{+}_{s}} dt ~ \J_{s}^{2}(z_{s})\nn \\
z_{s} &=& \frac{\xi}{\eta_{p}} \sqrt{2s\eta_{p}\, t - \left(1+\xi^{2}\right) t^{2} - \delta^{2}}, \label{eqn:rateCL}
\eea
where $\J_{s}$ is the $s$th-order Bessel function of the first kind, $\delta = m_{\phi}/m_e$, and $t = \eta_{k}/\eta_{p}$ is the \emph{lightfront fraction}, where
\[
t_{s}^{\pm} = \frac{s\eta_{p}}{1+\xi^{2}} \left(1 \pm \sqrt{1-\frac{\delta^{2}(1+\xi^{2})}{s^{2}\eta_{p}^{2}}}\right),
\]
and the threshold harmonic is ${s_{0}^\phi = \lceil (\delta\sqrt{1+\xi^2})/\eta_p \rceil}$, ($\lceil \cdot \rceil$ denotes the ceiling function). The rate is the number of scalars $N_{\phi}$ per unit phase duration, $L_\varphi$, in which the electron is in the electromagnetic wave. We take $L_\varphi=\varkappa^0\tau$ where $\tau$ is the duration of the wave. Expanding in $\xi \ll 1$ we find that the order-$s$ harmonic scales as $\xi^{2s}$, therefore the dominant contributions to the rate come from the $s=1$ contribution.
Performing the expansion of \eqnref{eqn:rateCL} for $\xi \ll 1$ we find the differential rate is
\bea
\frac{d\tsf{R}^{e\to\phi}}{dt}\simeq \frac{\xi^2g_{\phi e}^{2}}{16\pi \eta_p^3}\left(2\eta_pt-t^2-\delta^2\right) \qquad
\eea
where $t$ varies from $t_1^-$ to $t_1^+$, and the total rate becomes:
\bea
\tsf{R}^{e\to\phi}\simeq \frac{\xi^2g_{\phi e}^{2}}{12\pi}\left(1-\frac{\delta^2}{\eta_p^2}\right)^{3/2}. 
\eea

\subsection{Particle-In-Cell (PIC) code implementation}

\noindent Using the classical correspondence in \eqnref{eqn:swap} and the scalar wave equation from \eqnref{eqn:sys1}, scalar emission through nonlinear Thomson scattering can be included straightforwardly in numerical particle-in-cell codes. This simply requires using current methods for including standard low-energy nonlinear Thomson scattering from the vector current density, $j^\mu=\bar{\psi}\gamma^\mu\psi$, to be also applied to including the scalar current density $j=\bar{\psi}\psi$. It is important to note that the PIC codes only model radiation that can be resolved by the grid used in the numerical modelling.
For higher energy emission the results of these calculations break down and, as we will show, one must use the full QED result from \eqnref{FullQED}.

PIC codes are useful for capturing effects such as the coherent emission due to the presence of densely populated electron bunches.
However in Section \ref{coherence} we will demonstrate how these effects can also be included analytically.

\section{Comparison with QED result} \label{QEDcompare}

\noindent The QED result for this process, as calculated in \cite{king18b} can be written as $\tsf{R}^{e\to\phi}_{\tsf{QED}} = \sum_{s\geq s_{0}^\tsf{Q}}\tsf{R}^{e\to\phi}_{s,\tsf{QED}}$, where
\bea\label{FullQED}
\tsf{R}^{e\to\phi}_{s,\tsf{QED}} &=& \frac{g_{\phi e}^{2}}{16\pi \eta_{p}}\int_{u_{s}^{-}}^{u_{s}^{+}} \frac{du}{(1+u)^{2}}\{\left(4 - \delta^{2} \right)\J_{s}^{2}(z_{s}^{\tsf{Q}}) + \nn \\
&&\!\!\!\!\frac{\xi^{2}u^{2}}{2(1+u)} \left[ \J_{s+1}^{2}(z_{s}^{\tsf{Q}}) + \J_{s-1}^{2}(z_{s}^{\tsf{Q}}) - 2\J_{s}^{2}(z_{s}^{\tsf{Q}})\right]\},\nn \\ \label{eqn:Rphis}
\eea
where
\bea
\left(z^{\tsf{Q}}_{s}\right)^{2} = \left(\frac{2s\xi}{\sqrt{1+\xi^{2}}}\right)^{2} \frac{u}{u_{s}}\left(1-\frac{u}{u_{s}}\right) - \frac{\delta^{2}\xi^{2}(1+u)}{\eta_{p}^{2}},\nn \\ \label{eqn:zz}
\eea
and $ u= \eta_{k}/\eta_{q}$ with $\eta_{q} = \eta_{p}-\eta_{k}$ and $u_{s} = 2s \eta_{p}/(1+\xi^{2})$ with integration bounds $u_{s}^{\pm}$:
\bea
u_{s}^{\pm} = \frac{2s\eta_{p}-\delta^{2}}{2(1+\xi^{2})} \left[1 \pm \sqrt{1-\frac{4(1+\xi^{2})\delta^{2}}{(2s\eta_{p}-\delta^{2})^{2}}}\right]
\eea
and the threshold harmonic in the quantum case is ${s_{0}^\tsf{Q} =\lceil (2\delta\sqrt{1+\xi^2}+\delta^2)/2\eta_p\rceil}$.
This is the same threshold limit found in the classical case up to corrections of the order $\mathcal{O}(\delta^2)$.
The classical limit should correspond to the limit $\hbar \to 0$. 
Here we show how, when one takes this limit, we recover our classical expression. 
At this point we temporarily reinstate $\hbar$ and $c$ in the following paragraph.

The first thing to note about the QED calculation is the appearance of $\eta_{q}$, which is the energy parameter of the electron after it has emitted a photon. 
This parameter is absent in the classical description because EM radiation is not quantised, and therefore there is no recoil and the electron's energy parameter remains as $\eta_{p}$ during radiation of the EM field, which is continuous and not discrete.
This is clear from the fact that the photon energy parameter $\eta_{k} = \hbar^{2} \vkap\cdot k/mc^{2}$, is a power of $\hbar$ higher than the electron energy parameter $\eta_{p} = \hbar \vkap \cdot p/mc^{2}$. 
Therefore, in the classical limit $\eta_{q} \to \eta_{p}$ and so $u\to t$. 
Second, we note that $u \propto \hbar$ and $du/\eta_{p} \propto \hbar^{0} dt$, and so taking the limit of $\hbar\to 0$ of \eqnref{eqn:Rphis} gives:
\bea
\lim_{\hbar \to 0 }\tsf{R}^{e\to\phi}_{s,\tsf{QED}} &=& \frac{g_{\phi e}^{2}}{4\pi \eta_{p}}\int_{t_{s}^{-}}^{t_{s}^{+}} dt~\J_{s}^{2}(z_{s}).  \label{eqn:RphisB}
\eea
\eqnref{eqn:RphisB} is exactly the classical rate in \eqnref{eqn:rateCL}, which we arrived at using the ansatz \eqnrefs{eqn:swap}{eqn:swap2}. 
It is noteworthy that the mass term $\hbar^{2}\,k\cdot k = m_{\phi}^{2}c^{4}$ coefficient of the Bessel function disappears in the $\hbar \to 0$ limit, but the mass-term in the argument of Bessel function remains. 
This behaviour has also been observed by Erber and Latal \cite{0143-0807-24-1-308}, when they studied the correspondence between the quantum and classical results for radiation processes in a medium, where a non-zero index of refraction has a similar effect on the photon dynamics as a mass term does for the scalar field, and can be seen by integrating Eq. (11) in $\omega'$ in \cite{king16b} for the classical limit of nonlinear Compton scattering in a non-null transverse plane-wave EM background.  At this point we reset $\hbar=c=1$.
\begin{figure}[h!!] 
\centering
\includegraphics[draft=false,width=8.2cm]{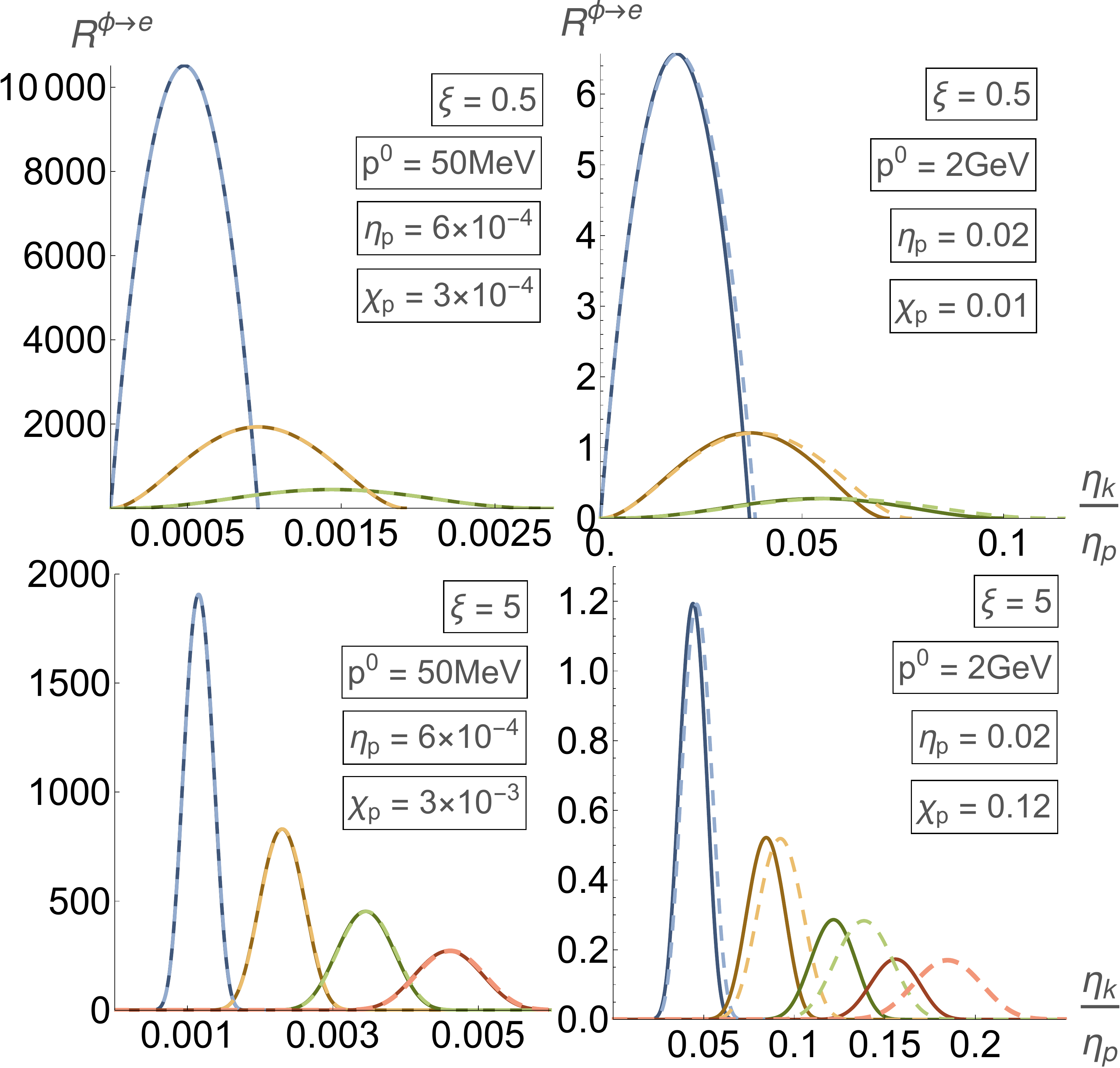}
\caption{
Here we compare the classical (dashed) and quantum (solid) spectra for scalar emission for a head-on collision of electron and laser-background, with $m_{\phi} = 1\,\trm{meV}$, $\vkap^{0} = 1.55\,\trm{eV}$ and $g_{\phi e} = 1$. In the first row the harmonics $s=1,2,3$ are shown and in the second row: $s=50,100,150,200$. (In comparison, tail-on collisions have $\eta_{p} \leq \vkap^{0} p^{0}/m^{2} \approx 3 \times 10^{-6}$.)}\label{fig:spec1}
\end{figure}

The accuracy of the classical limit can be ascertained by plotting the spectrum of emission of a single scalar by a single electron, which corresponds to comparing the integrands in \eqnrefs{eqn:rateCL}{eqn:Rphis}. We distinguish the perturbative ($\xi\ll1$) and all-order ($\xi\not\ll1$) cases for low and high-energy electron seeds, in \figref{fig:spec1}. It can be seen that in general for higher seed electron energy, the classical spectrum tends to predict a higher energy emitted per harmonic than the QED result (as in the comparison of nonlinear Thomson scattering to nonlinear Compton scattered photons \cite{dipiazza10}) and that at higher $\xi$, the discrepancy is larger. For the electron recoil from scalar emission to be negligible, and hence the classical limit to be a good approximation, the quantum nonlinearity parameter of the scalar: $\chi_{k} = \eta_{k}\xi$, must satisfy $\chi_{k} \ll 1$. This agrees with the comparison made in  \figref{fig:spec1}.

\begin{figure}[h!!] 
\centering
\includegraphics[draft=false,width=8cm]{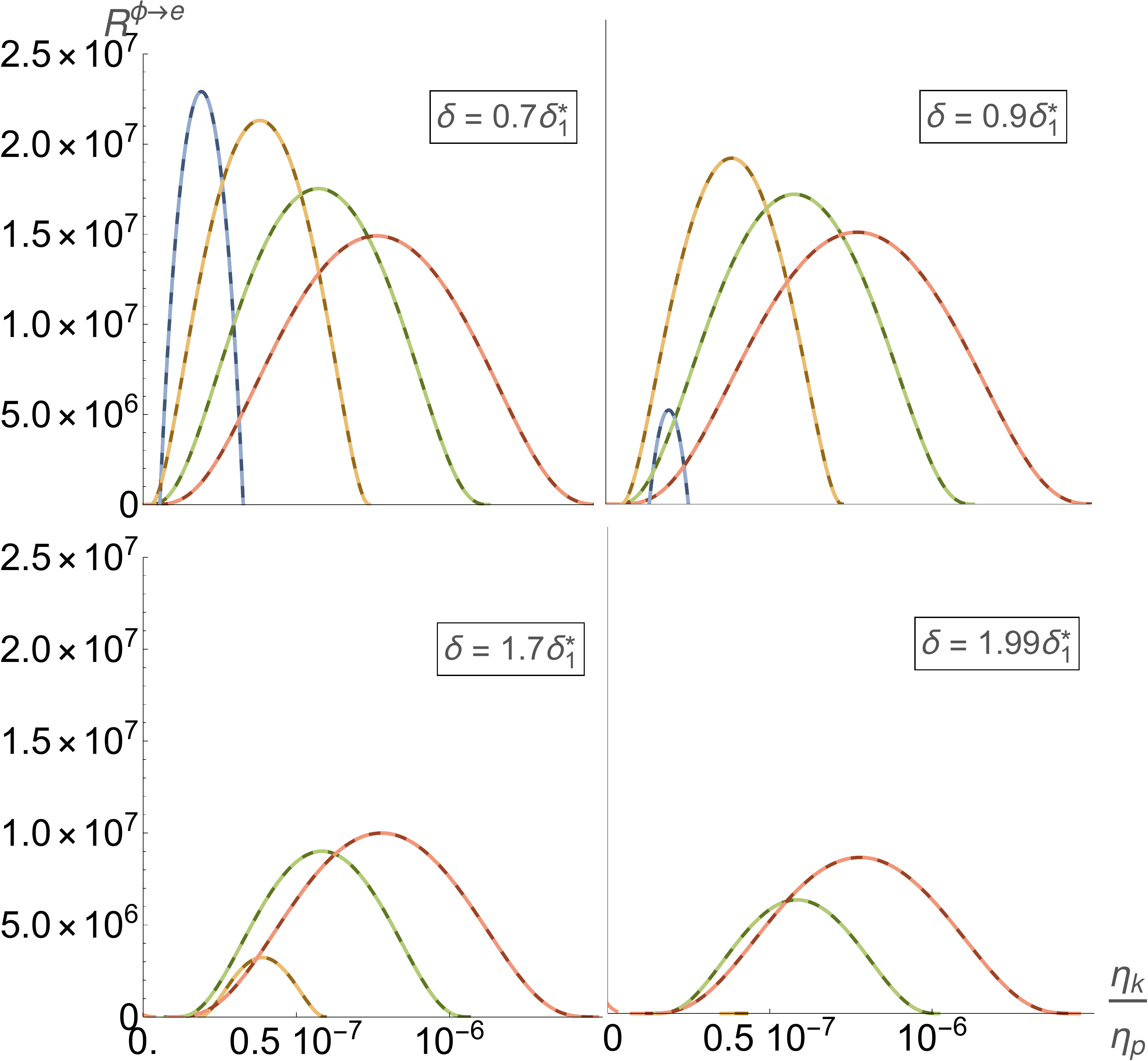}
\caption{The classical (dashed) and quantum (solid) spectra for a head-on collision of the laser background with $\xi=10$, $\vkap^{0} = 1.55\,\trm{eV}$ and an initial electron of energy of $p^{0} = 1.6\,\trm{MeV}$ ($\approx 2 \times 10^{-5}$).
For these parameters, $\delta_{1}^{\ast}=1\,\trm{eV}$. As the scalar mass is increased, the first and second harmonics are seen to disappear (each plot has the same axis scale).\label{fig:specmass}}
\end{figure}

In addition to comparing classical and quantum rates, we give a demonstration of the effect of the finite mass of the scalar. In \figref{fig:specmass} the value of the scalar mass is increased to show a ``channel-closing'' phenomenon. We define $\delta_{s}^{\ast}$ such that:
\[
 \delta = \delta_{s}^{\ast} \quad \Rightarrow \quad u_{s}^{+} = u_{s}^{-}.
\]
In other words, if $\delta >\delta_{s}^{\ast}$, the kinematic conditions required to emit the $s\,$th harmonic are forbidden. For the classical limit, this has a straightforward expression ${\delta_{s}^{\ast} = s\eta^{\ast}_{p}}$, where $\eta^{\ast}_{p} = \eta_{p}/\sqrt{1+\xi^{2}}$ is the energy parameter for an electron with an effective mass $m_{\ast} = m\sqrt{1+\xi^{2}}$.  Keeping $\eta_{p}$ and $\xi$ fixed, and considering different scalar masses, it can then be seen that if the scalar is massive enough, lower harmonics are suppressed. In \figref{fig:specmass} we choose parameters such that $\delta_{1}^{\ast}=1\,\trm{eV}$. At low electron energy, this effect is independent of whether the classical or quantum description is used.

\section{Coherent Emission}\label{coherence}

\noindent Coherent emission of radiation by electrons in a bunch of length $l$ is ensured for wavelengths $\lambda\gg l$, as there is no appreciable change in the phase of radiation emission over the bunch \cite{berryman96}. 
Decades ago, the FIREFLY experiment at the Stanford Linear Acceleration Center demonstrated that wavelengths even as short as $5\,\mu\trm{m}$ were emitted coherently from a $600\,\mu\trm{m}$ long electron bunch \cite{berryman96}. 
We can see this by considering the following scalar current for a bunch of $N_e$ electrons:
\bea
j(x^\prime)=g_{\phi e}\sum_{i=1}^{N_e} \int d\mathfrak{t}~\delta^{(4)}(x^{\prime\mu}-x^\mu (\mathfrak{t})-r^\mu_i).
\eea
The path $x^\mu (\mathfrak{t})$ denotes the centre of mass motion for the electron bunch and $r^\mu_i$ is the displacement of each electron from $x^\mu(\mathfrak{t})$. (In other words, the path of the $i$th electron is $x_{i}^\mu (\mathfrak{t}) = x^\mu (\mathfrak{t}) + r^\mu_i$). Taking the square of the Fourier transform we have
\bea
|\widetilde{\jmath}(k)|^2&=\Big|\sum_{i=1}^{N_e}e^{ik\cdot r_i}\Big|^2\Big|\widetilde{\jmath}_{1e}(k)\Big|^2=F(N_e,k,r)\Big|\widetilde{\jmath}_{1e}(k)\Big|^2 \nn \\
\eea
where $\widetilde{\jmath}_{1e}(k)$ is the Fourier transform of the one-electron current and the bunch effects are described by
\bea
F(N_e,k,r)=N_e+2\sum_{i=1}^{N_e-1}\sum_{j=i+1}^{N_e}\cos\left[k\cdot (r_i-r_j)\right]. \label{eqn:Ffac}
\eea
When the $k\cdot(r_i-r_j)$ factor is, or is close to, zero or a multiple of $\pi$, the effect of coherence on the production rate can be very large.
In an experimental set-up it is feasible to engineer the electron bunch and laser parameters such that $k\cdot(r_i-r_j)$ is close to zero. 
In \eqnref{eqn:Ffac} we see that if $\cos\left[k\cdot (r_i-r_j)\right] \to 1$, $F\to N_{e}^{2}$, if $\cos\left[k\cdot (r_i-r_j)\right] \to 0$, $F\to N_{e}$, but a random phase is approximated by using alternating signs with $\cos\left[k\cdot (r_i-r_j)\right] \to (-1)^{i-j}$, $F\to N_{e}\, \textrm{mod}\, 2$, representing destructive interference.

We will consider collimated bunches of electrons propagating in the $z$-direction such that $r^\mu=(0,0,0,r_z)^\mu$, and therefore $k\cdot r_i=-(k^++k^-)r_{zi}/2$ \footnote{Here we use lightfront coordinates where $p^{\pm}=p^0\pm p^3$, $p_{\pm}=2p^{\pm}$, and $p^\perp=(p^1,p^2)$.}.
To model an electron bunch we will take $N_e\sim 10^9-10^{10}$ electrons and choose their phases randomly from a Gaussian distribution with a standard deviation, $l$.
The term $r_{zi}$ measures the distance of the $i^{\text{th}}$ electron from the centre of the bunch.

\begin{figure}[h!!!] 
\centering
\includegraphics[draft=false,width=8cm]{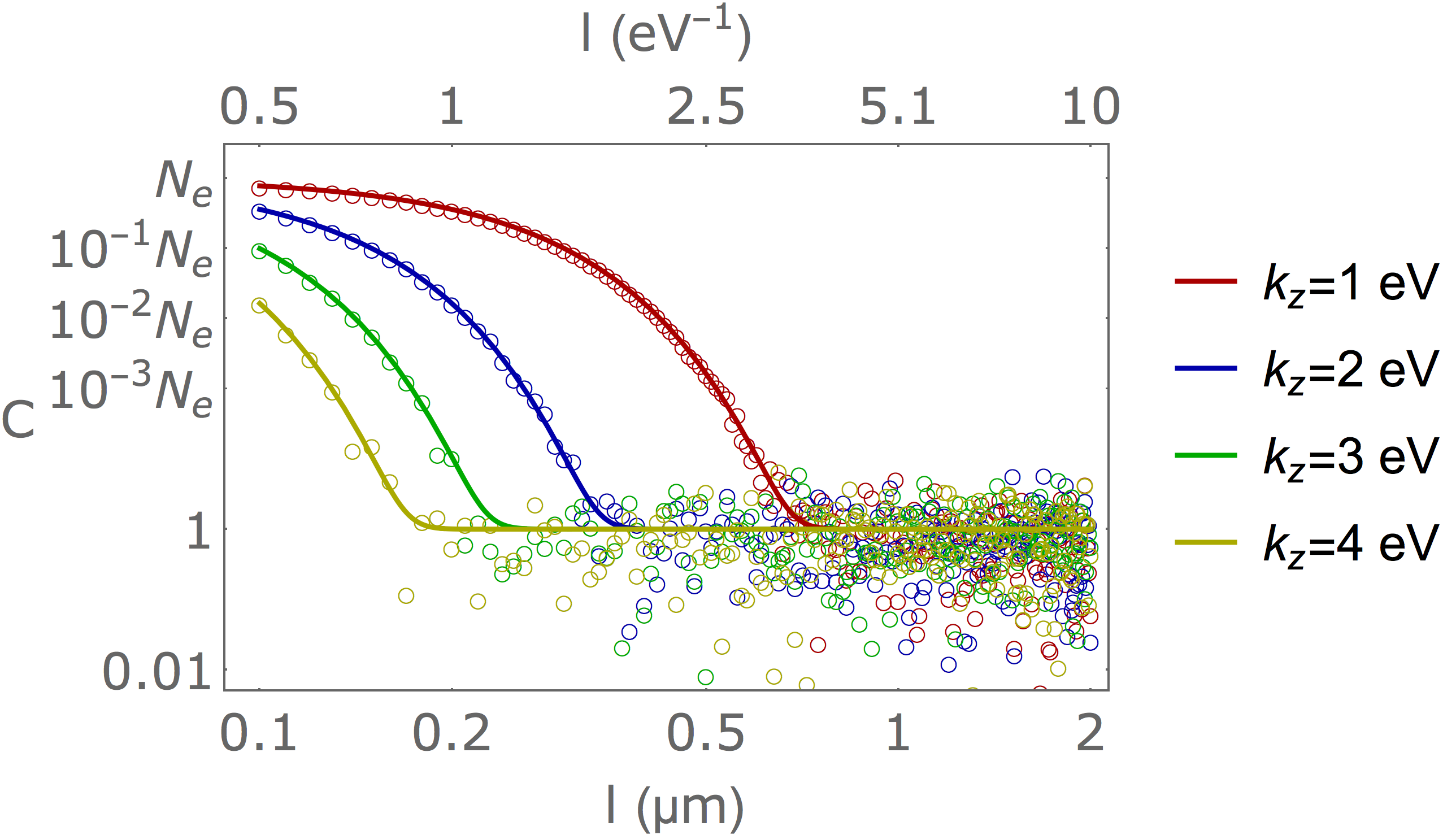}
\caption{The coherence changes as one varies the bunch length of a Gaussian-shaped bunch of $10^5$ electrons,
where $|\widetilde{\jmath}(k)|^2=\mathcal{C}N_e|\widetilde{\jmath}_{1e}(k)|^2$. 
The empty circles are the numbers generated from the random Gaussian distribution, and the solid lines are given by the approximating function $\mathcal{C}_*$.
}\label{FigCoherence}
\end{figure}

We define the coherence factor $\mathcal{C}=F/N_e$, and with the electrons distributed just in the $z$ direction we have $k\cdot (r_i - r_j)=-k_z(r_{zi}-r_{zj})$.
Choosing $r_{zi}$ and $r_{zj}$ from a Gaussian distribution with a standard deviation equal to $l$ results in the coherence factor being well approximated by the function $\mathcal{C}_*=1+N_ee^{-\sigma_*^2}$ where $\sigma_*=k_zl$.
Therefore, coherence effects are important when $\sigma_*\ll \sqrt{\ln(N_e)}$.
In Fig. \ref{FigCoherence} we plot the size of the coherence effects as a function of the bunch length for various values of $k_z$.

Including the coherence effects leads to a modification in the yield presented in  \eqnref{yield1},
\bea\label{yieldCoh}
N_{\phi} = \int \frac{d^{3}k}{(2\pi)^{3}}\frac{1}{2k^{0}} \left(1+N_ee^{-\sigma_*^2(k)}\right)|\widetilde{\jmath}_{1e}(k)|^{2},
\eea
where we recall that $\widetilde{\jmath}_{1e}(k)$ is the Fourier transform of the one-electron current.
The differential distribution $d^3N_{\phi}/dk^3$ will then have a coherence enhancement towards the lower end of the spectrum.

The coherence properties are explicitly dependent on the polar angle at which the scalar is emitted, where $k_z=|\vec{k}|\cos\theta$ with $\theta$ being the polar angle with the positive $z$ axis.
So it is instructive to use spherical polar coordinates rather than lightfront coordinates to study the coherence effects, and this is also useful when considering an experimental set-up to detect these scalar particles.
Using spherical polar coordinates the classical result for the total rate can be written as
\bea
\begin{split}
\tsf{R}^{e\to\phi}=&\frac{g_{\phi e}^2}{(2\pi)^3}\frac{1}{(\kappa^0)^2}\frac{N_e}{2}\sum_{s=1}^{\infty}\sum_{n=1}^2\int d|\vec{k}|d\theta d\phi~\frac{|\vec{k}|^2}{k^0}\sin\theta \\
& \frac{\delta(|\vec{k}|-|\vec{k}|_n)}{|w_0'(|\vec{k}|_{n})|}\left(1+N_ee^{-l^2|\vec{k}|^2\cos^2\theta}\right)\J_s^2\left(z_{s}\right),
\end{split}
\eea
where $n\in\{1,2\}$ tracks the two solutions for $|\vec{k}|$ in applying the global momentum-conserving delta-function and
\bea
w_0(|\vec{k}|)=\frac{1}{\varkappa\cdot p}\left(p\cdot k+\frac{(m_e\xi)^2}{2}\frac{\varkappa\cdot k}{\varkappa\cdot p}\right),
\eea
with $k^0=\sqrt{|\vec{k}|^2+m_\phi^2}$, ${k^1=|\vec{k}|\sin\theta\cos\phi}$, ${k^2=|\vec{k}|\sin\theta\sin\phi}$, and ${k^3=|\vec{k}|\cos\theta}$.
The solutions for $|\vec{k}|_i$ are obtained by solving $w_0-s=0$ and we can write the argument of the Bessel function as
\bea
\begin{split}
z^2=\left(\frac{\xi}{\eta_p}\right)^2\Big[2\eta_ks-\delta^2-\left(\tfrac{\eta_k}{\eta_p}\right)^2(1+\xi^2)\Big]
\end{split}
\eea
with $\eta_k=\varkappa\cdot k/m_e^2=\varkappa^0(k^0-|\vec{k}|\cos\theta)/m_e^2$.
The rate depends explicitly on both the polar angle $\theta$ and the azimuthal angle $\phi$, with the latter dependence arising from evaluating the delta-function ($|\vec{k}|_n$ depends on $\phi$ in general).
However when $\epsilon\cdot p=\beta\cdot p=0$ the rate becomes independent of the azimuthal angle.
We will parametrise the incoming electron momenta with $p^0=\sqrt{|\vec{p}|^2+m_e^2}$, $p^1=|\vec{p}|\sin\theta_{p}\cos\phi_{p}$, $p^2=|\vec{p}|\sin\theta_{p}\sin\phi_{p}$, and $p^3=|\vec{p}|\cos\theta_{p}$.
The azimuthal dependency of the differential rate is trivially related to the azimuthal angle of the incoming electrons, so we will set $\phi_p=0$ to simplify interpretation of our results and hence describe the incoming electrons by their polar angle $\theta_p$ and their gamma factor $\gamma_p^{2}=1+|\vec{p}|^2/m_e^2$.

\begin{figure}[h!!!] 
\centering
\includegraphics[draft=false,width=6cm]{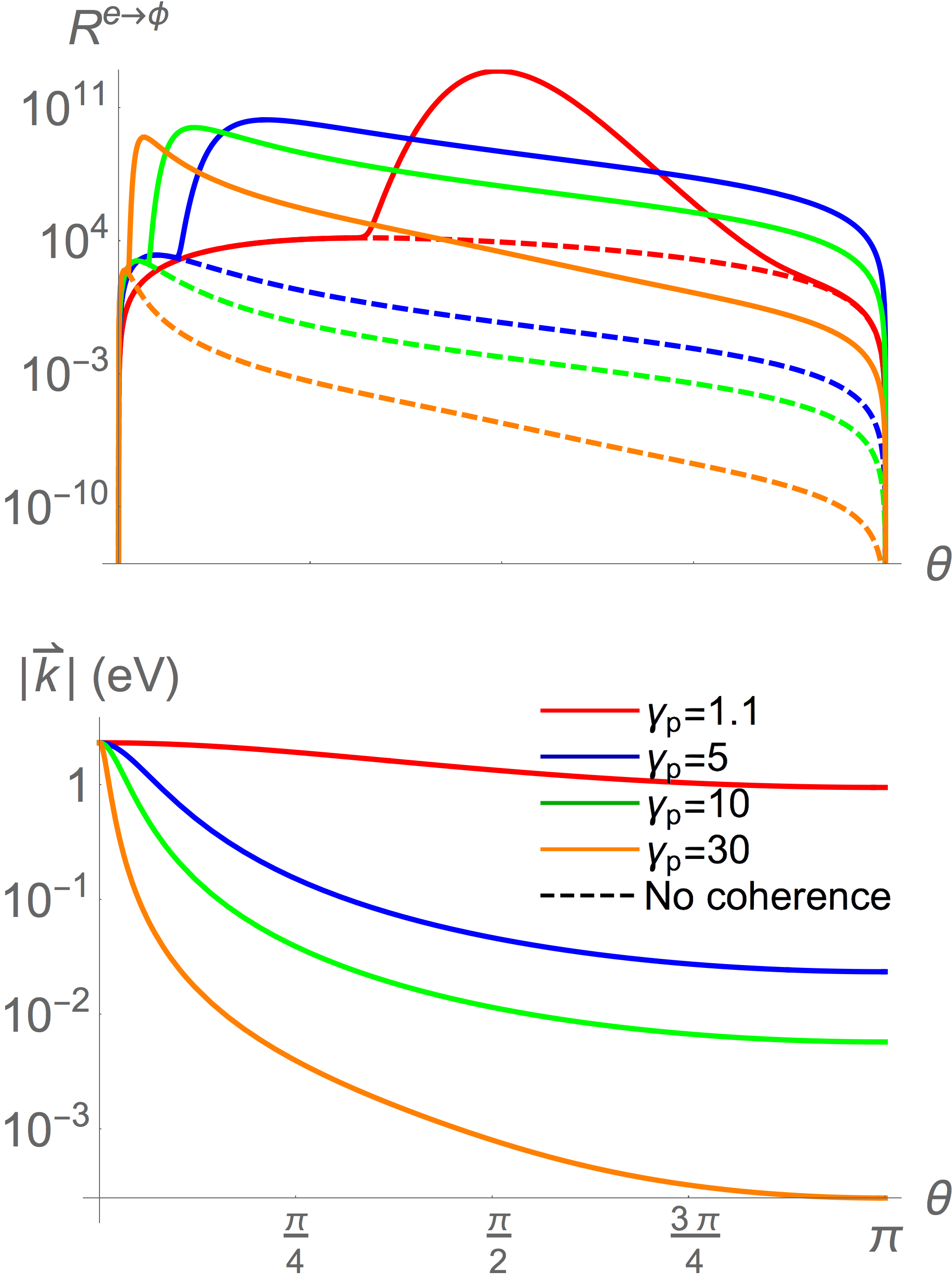}
\caption{In the upper plot we have the total rate as a function of the polar angle of the emitted scalar particles, and in the lower plot we have the emitted scalar momentum as a function of the polar angle.  We have assumed that $\theta_p=0$ such that the electrons and the colliding photons are co-propagating, and that the $s=1$ contribution dominates.
We have also taken $g_{\phi e}=1$, $N_e=10^9$, $\varkappa^0=2.33$ eV, $l=1~\mu$m, $m_\phi=1$ meV, and $\xi=0.1$ in this calculation.}\label{gridPlot1}
\end{figure}

We will start with the case that $\epsilon\cdot p=\beta\cdot p=0$ and the electrons and laser beam are co-propagating i.e. ``tail-on'' ($\theta_p=0$), where there is no azimuthal dependency.
In Fig. \ref{gridPlot1} we show how the total rate and the emitted scalar momentum depends on the polar angle of the emitted scalar.
We see that for small values of $\gamma_p$ the coherence effects are focused at emission angles $\theta\sim\pi/2$.
This is because at $\theta=\pi/2$ the coherence effects are maximised by a minimisation of $\sigma_*\sim\cos^2\theta$. A physical way of thinking of this is that the ``transverse'' bunch length is much smaller than the ``longitudinal'' beam length, so coherence effects are most pronounced when scalars are emitted transversally.
At larger values of $\gamma_p$ the peak at which coherence effects are focused is shifted towards smaller values of $\theta$ due to the well-known narrowing of the relativistic $\theta \sim 1/\gamma_{p}$ emission cone \cite{jackson99}. So one sees how the coherence enhancement at right-angles to the collision axis and the relativistic enhancement at small angles, combine to give a peak which moves from being perpendicular to the collision axis to being more along it, the more relativistic the incoming electrons are.
The coherence effects are sustained at $\theta>\pi/2$ and suppressed at $\theta<\pi/2$.
This can be understood from looking at the lower plot in Fig. \ref{gridPlot1} where we see that at $\theta>\pi/2$ the values of $|\vec{k}|$ are smaller and thus $\sigma_*$ is smaller.

\begin{figure}[h!!!] 
\centering
\includegraphics[draft=false,width=6cm]{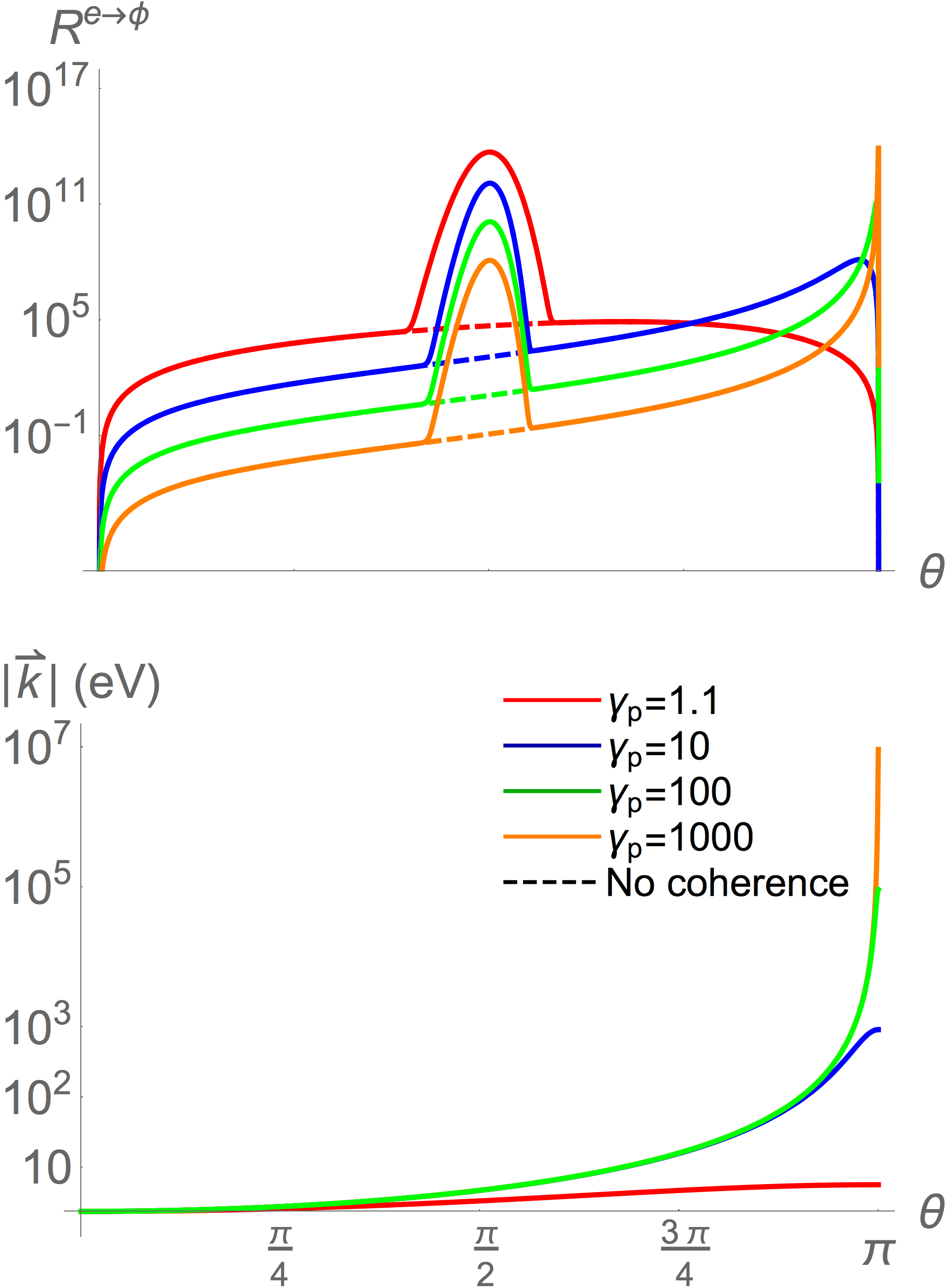}
\caption{In the upper plot we have the total rate as a function of the polar angle of the emitted scalar particles, and in the lower plot we have the emitted scalar momentum as a function of the polar angle.  We have assumed that $\theta_p=\pi$ such that the electrons and the colliding photons are counter-propagating, and that the $s=1$ contribution dominates.
In this plot, we also have $g_{\phi e}=1$, $N_e=10^9$, $\varkappa^0=2.33$ eV, $l=1~\mu$m, $m_\phi=1$ meV, and $\xi=0.1$.
}\label{gridPlot2}
\end{figure}

In Fig. \ref{gridPlot2} we have essentially the same information as in Fig. \ref{gridPlot1} except with the electrons counter-propagating or ``head-on'' to the laser beam, i.e. $\theta_p=\pi$.
Here we see that the coherence effects are completely lost for electron bunches with large $\gamma_p$ factors, this being due to the fact that the emitted scalars have much larger momenta than in the tail-on case and thus $\sigma_*$ is larger and the coherence effects more suppressed.
In addition to this, the coherence effects for the electron bunches with lower $\gamma_p$ are localised at $\theta\sim\pi/2$, again this is because this is the only parameter range at which $\sigma_*$ is small.
It is worth noting that our assumption of a collimated electron bunch is important here.
If we had a sizeable bunch width, then $\sigma_*$ would have an appreciable dependence on the azimuthal angle, and the enhancements at $\theta\sim\pi/2$ would be smoothed out.

It is important to note that in this analysis we have neglected electron-electron interactions. We first justify this with reference to recent laser-electron collision experiments, which demonstrate how that electrons can be accelerated from gas jets into bunches of length $\sim 10\,\mu\trm{m}$ and overlapped with the laser focus in the collision point \cite{sarri17,cole18}. Second, it can be shown that the force on the electrons due to the laser field is much stronger than the Coulomb repulsion between electrons in the bunches we have considered here, and neglecting this extra force is in line with other approximations we have made, such as neglecting radiation reaction \cite{angioi18}.

To detect the scalar particles emitted in these laser-electron interactions it is beneficial to have the majority of the emission in a small solid angle, such as along the collision axis, i.e. at $\theta\simeq 0$ or $\theta\simeq\pi$.
The experimental set-ups that most easily result in this scenario are those involving electron bunches with large $\gamma_p$.
We consider two scenarios:
\begin{itemize}
\item {\it tail-on collision}:  with $\gamma_p\gtrsim30$ approximately all of the scalars are emitted in the $0\leq\theta\lesssim0.1$ region, and coherence effects can drastically increase angular rates.
\item {\it head-on collision}:  with $\gamma_p\gtrsim300$ approximately all of the scalars are emitted in the $3.1\lesssim\theta\leq\pi$ region, however coherence effects are negligible for all scalar masses in this case.
\end{itemize}

Focusing on these two scenarios we will use the yields derived in terms of the lightfront momentum in Sec. \ref{NLTcalc} and \ref{QEDcompare}, with the inclusion of the coherence effects in Sec. \ref{coherence}.
We can estimate the energy of the scalar particles from the $\eta_k$ distribution assuming $\theta\simeq0\text{ or }\pi$,
\bea\label{ALPenergy}
E_\phi\simeq \frac{m_e^2}{\varkappa^0}\frac{\eta_k}{2}\left(1+\left(\frac{\varkappa^0\delta}{m_e\eta_k}\right)^2\right). \label{eqn:Ephi1}
\eea

\section{Experimental prospects for scalar ALP production and detection}\label{ExpProspects}

\noindent Several high power laser facilities now have the capability to produce intense laser pulses with $\xi$ of the order of $0.1$ to $1$ at a repetition rate of $1\,\trm{Hz}$, such as VEGA \cite{vega_site},
BELLA \cite{bella_site}, Draco \cite{Draco} and the the upcoming ELI-Beamlines laser facility \cite{elibeamlines}. Through collisions with fixed targets these pulses can be used to produce high energy ($\mathcal{O}(\text{GeV})$) electron bunches with $N_e\sim10^9$ and $l=\mathcal{O}($10$)\,\mu$m \cite{sarri15}.

In this section we propose an outline for the first lab-based experiment to probe the product of couplings $g_{\phi e} g_{\phi \gamma \gamma}$. The set-up we envisage is similar to that of LSW set-ups, but where in a generation region, an electron beam collides with a laser pulse to produce massive scalars, and in a regeneration region, which is shielded from the background produced in the generation region by a wall, massive scalars are converted into photons in the presence of a static magnetic field, which are measured in this low-noise region. Many experiments already use similar techniques to search for light scalar and pseudoscalar particles in lab-based environments, for example the ALPS experiment \cite{ALPS} (and its upcoming successor \cite{Bahre:2013ywa}, as well as other planned experiments such as STAX \cite{Capparelli:2015mxa}).
The CAST experiment uses the same detection technique to search for axions produced in the Sun \cite{CAST17}.

In the generation region, laser pulses from the facilities mentioned above can be split such that one pulse collides with a fixed target producing a bunch of electrons, while the other pulse collides with the bunch of electrons. This allows the two set-ups: ``tail-on'' and ``head-on', to be realised.

In the regeneration region, we envisage a strong magnetic field (strength $B$) extending over some length $L$, in which the massive scalars decay to a photon through the coupling described in the introduction,  $\mathcal{L}^{I}_{\phi\gamma\gamma}=-g_{\phi\gamma\gamma}\phi F^{\mu\nu}_{B}F_{B\,\mu\nu}$. Then in contrast to \eqnref{eqn:sys1}, the system of equations in the \emph{regeneration} region, is:
\bea
\left(\Box + m^{2}_{\phi}\right)\phi &=& -g_{\phi \gamma\gamma}\tr F^{2} \nn \\
\left(1+4g_{\phi\gamma\gamma}\phi\right)\Box A^{\nu} &=& -4g_{\phi \gamma \gamma} F^{\mu\nu}\partial_{\mu}\phi. \label{eqn:sys2}
\eea
Again, making a substitution $F \to F_{B} + F_{\gamma}$, and a perturbative ansatz in $F_{\gamma}$, we have, to lowest-order in $g_{\phi\gamma\gamma}$:
\bea
\left(\Box + m^{2}_{\phi}\right)\phi &=& -g_{\phi \gamma\gamma}\tr F_{\gamma}^{2} \label{eqn:sys2einfach},
\eea
where $\Box A_{B} = 0$, $\Box A_{\gamma} = 0$. The detection of photons in the low-noise regeneration region is then the  experimental signal.

As a benchmark to evaluate the effectiveness of our proposed set-up we will assume a laser pulse with $\xi=0.1$, $\varkappa^0=2.33\,$eV, and a repetition rate of $1\,$Hz, collides with an electron bunch of electrons with initial energies ranging from MeV to tens of GeV. For the detection region, we assume the same parameters as in the ALPS experiment: a $B=5\,$T magnet which extends over $L=4.21\,$m and photon detectors with a dark count rate of $n_b=10^{-4}\text{s}^{-1}$.

Adopting the benchmarks set out at the end of Sec. \ref{coherence} we can assume that all the produced scalar particles enter the regeneration region at approximately $\theta=0$ or $\theta=\pi$.
In this case, the probability of the scalar particle decaying to a photon in the magnetic field is:
\bea
P_{\phi\rightarrow\gamma}=&\left[2\,\dfrac{g_{\phi\gamma\gamma} B E_\phi}{m_\phi^2}\sin\left(\dfrac{m_\phi^2L}{4E_\phi}\right)\right]^{2}
\eea
where $E_\phi$ is the energy of the scalars entering the detection region (see e.g. \eqnref{eqn:Ephi1}).
The probability of regeneration extends to larger masses for larger scalar energies.
For $E_\phi\gg m_\phi^2L/4$ we have $P_{\phi\rightarrow\gamma}\simeq (g_{\phi\gamma\gamma}BL/2)^{2}$ and the probability is enhanced by the extent and strength of the magnetic field.

\section{Exclusion bounds} \label{ExclBounds}

\noindent For the benchmarks defined at the end of Sec. \ref{coherence} we can write the total number of photons, $N_{\gamma}$, converted from scalars, per electron-laser-pulse collision as
\bea
N_{\gamma}&\simeq& N_eL_{\vphi} \int_{t_{1}^{-}}^{t_{1}^{+}} dt~\frac{d\tsf{R}_{e\rightarrow\gamma}}{dt}P_{\phi\rightarrow\gamma} \nn\\
&=&4N_eL_{\vphi}\frac{g_{\phi\gamma\gamma}^2B^2}{m_\phi^4}\int_{t_{1}^{-}}^{t_{1}^{+}} dt~\frac{d\tsf{R}_{e\rightarrow\gamma}}{dt}E_\phi(t)^2\sin^2\left(\frac{m_\phi^2L}{4E_\phi(t)}\right).\nn\\
\eea
In the $\xi \ll 1$ and $m_\phi^2L /4  \ll E_\phi$ limit, neglecting coherence effects, this simplifies to
\bea
N_{\gamma}\simeq N_e\frac{\xi^2L_\varphi g_{\phi\gamma\gamma}^2g_{\phi e}^2}{48\pi}(BL)^2\left(1-\frac{\delta^2}{\eta_p^2}\right)^{3/2}.
\eea
When coherence effects dominate we would find $N_{\gamma}\sim N_e^2$ rather than $N_{\gamma}\sim N_e$.
Outside the $m_\phi^2L \ll E_\phi$ region $N_\gamma$ scales as $m_\phi^4$ and therefore the bounds are less restrictive.

\begin{figure} 
\centering
\includegraphics[draft=false,width=8.5cm]{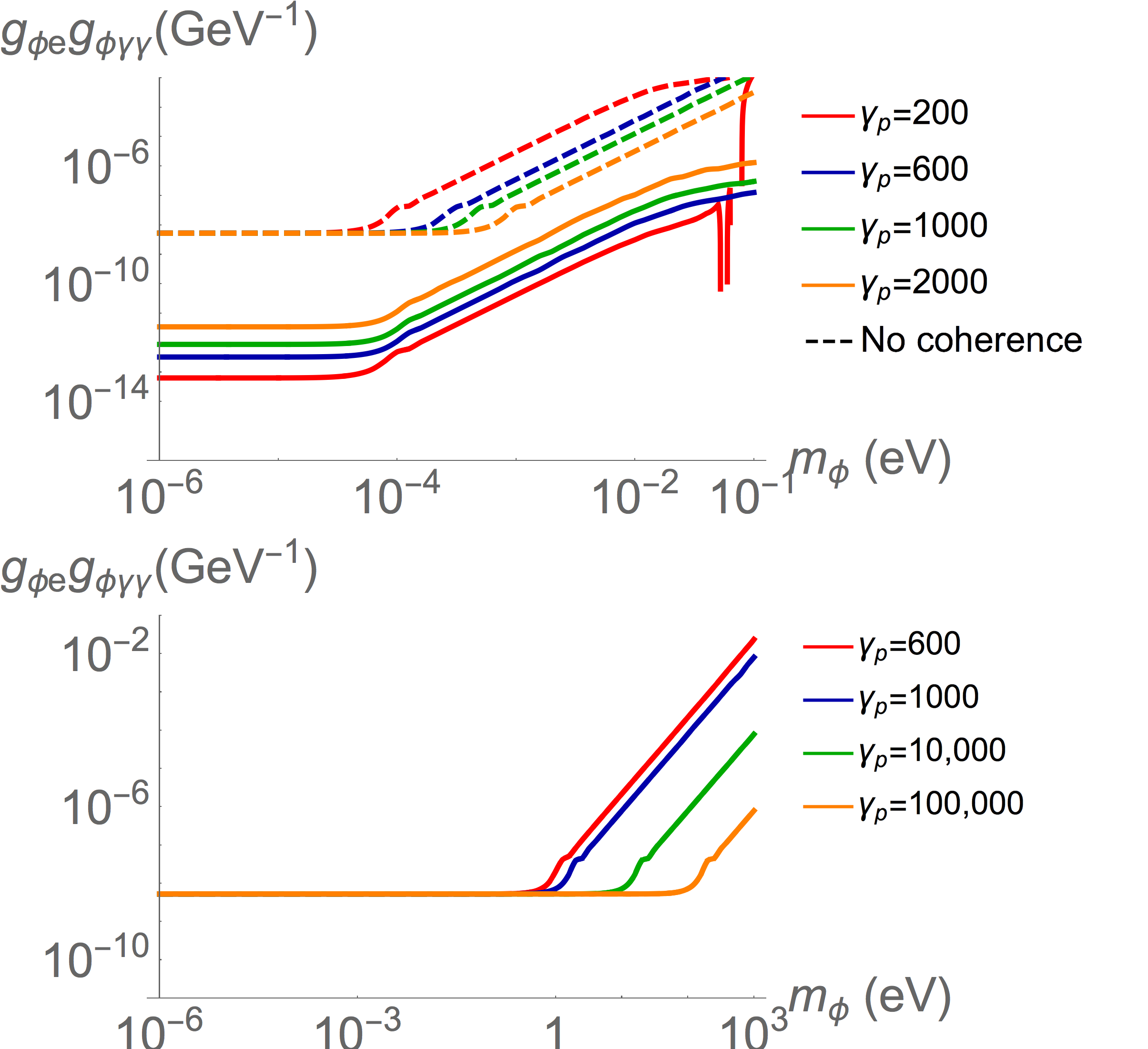}
\caption{Projected exclusion bounds from the proposed experimental set-up with tail-on (top) and head-on (bottom) collisions between the electron bunch and laser pulse with an ALPS I-like detection region.  The regions above the lines would be excluded, and the coherence effects are only relevant for the tail-on collisions.
}\label{exclPlot}
\end{figure}

We assume that the laser pulses have duration $\tau= 100$ fs, are of intensity parameter $\xi\simeq0.1$, and are produced at a rate of $1\,$Hz.
We assume that each pulse collides with a bunch of $10^{10}$ electrons of length $l=10\,\mu$m and that the experiment runs for a total of $100$ hours.
The projected bounds from such an experiment with various electron $\gamma_p$ factors are shown in Fig. \ref{exclPlot} for both tail-on and head-on collisions.
In deriving these projected bounds we have used the full expressions for the differential yield, including coherence effects, and evaluated the expressions numerically.

The first plot in Fig. \ref{exclPlot} shows the projected exclusion bounds for a tail-on collision, where we see that the coherence effects are significant for all scalar masses, and that a degradation effect scaling as $m_\phi^{-4}$ begins at ${m_\phi\sim0.1\,\trm{meV}}$.
Increasing the $\gamma_p$ factor of the incoming electrons does not significantly affect the point at which this degradation occurs, and in fact only suppresses the coherence effects.
In the second plot we see the projected exclusion bounds for a head-on collision, where the coherence effects are entirely negligible.
In this case however, the scale at which the $m_\phi^{-4}$ degradation occurs is significantly affected by the $\gamma_p$ factor of the incoming electrons.
For $\gamma_p=10,000$ and $100,000$ (electron energies of $\simeq 5$ GeV and $\simeq 50$ GeV), the scale at which degradation occurs is pushed to $m_\phi\simeq10$ eV and $100$ eV, respectively.

Through the coupling of a scalar field to the electromagnetic field, a coupling between the scalar field and nucleons is induced at one-loop order.
The scalar-nucleon coupling is severely constrained by both astrophysical and lab-based fifth force experiments \cite{Dupays:2006dp}.
The ALP-photon coupling arises through a dimension-five operator in the Lagrangian, and if one assumes that long-distance effects occur at scales much larger than $m_\phi$ then the constraints from fifth force experiments imply that $g_{\phi\gamma\gamma}(\text{GeV}^{-1})\lesssim 10^{-10}$ for $m_\phi\sim0.1\,$eV and $g_{\phi\gamma\gamma}(\text{GeV}^{-1})\lesssim 10^{-17}$ for $m_\phi\sim10^{-6}\,$eV.
The CAST experiment also places a similarly strong bound on the product of couplings $g_{\phi e}g_{\phi\gamma\gamma}(\text{GeV}^{-1})\lesssim 10^{-22}$ for $m_\phi\lesssim10^{-2}\,$eV, and the degradation of this bound for heavier masses scales as $\sim m_\phi^{-4}$ \cite{Barth:2013sma}. 
The heaviest masses probed by the CAST experiment were $m_\phi\sim1.2\,\trm{eV}$.
When the PVLAS experiment reported a signal contradicting these bounds there were models proposed which partially evaded these astrophysical and fifth force constraints by reducing the bounds by several orders of magnitude \cite{Masso:2005ym,Jain:2005nh,Jaeckel:2006id,Masso:2006gc}.
Despite the fact that this result has since vanished, the need for lab-based tests of light ALPs coupled to photons and electrons is apparent.

The most recent results from these experiments imply a bound $g_{\phi\gamma\gamma}(\text{GeV}^{-1})\lesssim 10^{-7}$ for $m_\phi\sim10^{-3}\,$eV and the degradation of this bound for heavier masses scales as $\sim m_\phi^{-8}$.
These bounds from the ALPS I experiment are the most sophisticated lab-based bounds available for light scalar particles.
Given the projected exclusion bounds presented in Fig. \ref{exclPlot} we conclude that the experimental set-up proposed in the current paper would provide an excellent complementary set of lab-based bounds on the parameter space of light scalar particles.
The benefits here are two-fold: from the tail-on collisions one is able to obtain a high precision on the ALP-photon and ALP-electron couplings at ${m_\phi\lesssim10^{-4}\,\trm{eV}}$, and from the head-on collisions one is able to push the mass range over which these experiments are sensitive to ${m_\phi\lesssim 100\,\trm{eV}}$. This could be achieved, for example, by using the $17.5\,\trm{GeV}$ electron beam from the XFEL at DESY and combining it with an ALPS-style dipole magnet. These results could be significantly improved by better technology on the production side of the experiment where the scalar particles are produced in laser-electron collisions, i.e. through larger repetition rates, denser electron bunches, or longer run times.

\section{Conclusion}

\noindent We started by demonstrating the equivalence between classical and quantum emission of scalar particles via non-linear Compton scattering in interactions between an electron and an intense laser in the classical $\hbar\rightarrow0$ limit (equivalently the disappearing lightfront momentum limit $\eta_{p}\to 0$), and detailed how these processes can be included in PIC code simulations.
(For a discussion on the pseudoscalar case see Appendix \ref{AppA}.)
We then looked at possible coherence effects due to the dense population of the electrons in the collision with the laser pulse.
It is evident that collisions in which the laser pulse collides with the incoming electrons while travelling in the same direction (i.e. $\theta_{p}\simeq0$, or `tail-on') result in the largest coherence effects, while `head-on' collisions (i.e. $\theta_{p}\simeq \pi$) only result in sizeable coherence effects for incoming electrons with small $\gamma_p$ factors.
In experiments designed to produce and detect exotic scalar particles in the lab, it is beneficial for the scalar particles to be produced in a collimated `beam', i.e. $\theta\simeq0$ or $\pi$.
We identified two scenarios in which this occurs, one is tail-on collisions with $\gamma_p\gtrsim30$, and the other is head-on collisions with $\gamma_p\gtrsim300$.
An example experimental set-up was discussed that had the ability to produce scalar particles through laser-electron interactions and detected through the conversion of the scalar particle to a photon in an external magnetic field.
Assuming the same detection technology present in the ALPS I experiment, projected exclusion bounds on the product of the $g_{\phi e}$ and $g_{\phi\gamma\gamma}$ couplings were computed.
In the tail-on collisions we have shown that bounds could be obtained on $g_{\phi e}g_{\phi\gamma\gamma}(\text{GeV}^{-1})$ of the order $10^{-13}$ for scalar masses below $\sim10^{-1}$ meV.
These bounds are not competitive with the bounds set by CAST or the fifth force experiments, but as explained in the text, those are model-dependent bounds that may be evaded in certain theoretical models.
In the head-on collisions we have shown that there are opportunities to probe scalar masses in range $10-100$ eV, outside the bounds derived from the CAST and fifth force experiments.
Beam dump experiments also place experimental bounds on the size of the ALP-electron coupling.
In \cite{Andreas:2010ms} bounds on the coupling of a pseudo-scalar ALP to electrons was obtained from data collected at previous flavour, reactor, and beam dump experiments.
These upper bounds are typically of the order $g_{\phi e}\lesssim10^{-4}-10^{-3}$.
A recent study has also analysed the bounds on the ALP couplings that could be obtained from the proposed LDMX experiment \cite{Berlin:2018bsc}.

Therefore to conclude, the experimental set-up suggested could indeed probe interesting regions of parameter space not yet studied in a completely lab-based environment, and it could provide very useful complementary bounds to those obtained in other lab-based LSW experiments, such as ALPS.

\section{Acknowledgments}
The authors acknowledge useful discussions with C. D. Murphy, A. Ringwald, and T. Heinzl. B. K. and B. M. D. acknowledge funding from Grant No. EP/P005217/1.

\appendix

\section{Pseudo-scalar production rate in the classical limit}\label{AppA}

\noindent In addition to measuring massive scalars, there is also much interest in measuring massive pseudoscalars - particularly as a partial solution to the dark matter question. Pseudoscalar creation from an electron in a laser pulse was studied in \cite{king18b,king18c}, where it was found that in the low-$\eta_{p}$ limit, the rate was heavily suppressed. The disappearance of the rate for pseudoscalar creation at low seed-particle energies can be understood through the classical limit. In the Weyl basis, the interaction $\phi \psibar \gamma^{5} \psi = \phi\left[\psibar_{L}\psi_{L}-\psibar_{R}\psi_{R}\right]$, and since classically, there is no difference between left-handed and right-handed electrons, it is consistent that the rate for pseudoscalar creation should be identically zero. The QED pseudoscalar rate can be arrived at from the QED scalar rate \eqnref{eqn:Rphis} by the replacement 
\[
4-\delta^{2} \to -\widetilde{\delta}^{2},
\]
where $\widetilde{\delta} = m_{\vphi}/m$ and $m_\vphi$ is the mass of the pseudoscalar. Just as in the massive scalar case, this term must disappear, and hence the $\hbar\to 0$ limit is indeed identically zero.

\bibliographystyle{elsarticle-num}
\bibliography{current}

\end{document}